\documentclass{aa}
\input epsf.sty
\begin{document}

\title{Optical and radio variability of the BL Lac object \object{AO
0235+16}: a possible 5--6 year periodicity}


\author{C.M.~Raiteri \inst{1}
\and M.~Villata \inst{1}
\and H.D.~Aller \inst{2}
\and M.F.~Aller \inst{2}
\and J.~Heidt \inst{3} 
\and O.M.~Kurtanidze \inst{4,5}
\and L.~Lanteri \inst{1}
\and M.~Maesano \inst{6}
\and E.~Massaro \inst{6}
\and F.~Montagni \inst{6}
\and R.~Nesci \inst{6}
\and K.~Nilsson \inst{7}
\and M.G.~Nikolashvili \inst{4}
\and P.~Nurmi \inst{7}
\and L.~Ostorero \inst{8}
\and T.~Pursimo \inst{7}
\and R.~Rekola \inst{7}
\and A.~Sillanp\"a\"a \inst{7}
\and L.O.~Takalo \inst{7}
\and H.~Ter\"asranta \inst{9}
\and G.~Tosti \inst{10}
\and T.J.~Balonek \inst{11} 
\and M.~Feldt \inst{12}
\and E.J.~McGrath \inst{11}
\and A.~Heines \inst{3}
\and C.~Heisler \inst{13} \thanks{Charlene passed away on October 28, 1999}
\and J.~Hu \inst{14}
\and M.~Kidger \inst{15}
\and J.R.~Mattox \inst{16}
\and A.~Pati \inst{17}
\and R.~Robb \inst{18}
\and A.C.~Sadun \inst{19}
\and P.~Shastri \inst{17}
\and S.J.~Wagner \inst{3}
\and J.~Wei \inst{14}
\and X.~Wu \inst{14}
}

\offprints{C.M.\ Raiteri, \email{raiteri@to.astro.it}}

\institute{Osservatorio Astronomico di Torino, Via Osservatorio 20, 
I-10025 Pino Torinese, Italy 
\and Dept.\ of Astronomy, Dennison Bldg., U.\ Michigan, Ann Arbor, MI
48109, USA
\and Landessternwarte K\"onigstuhl, W-6900 Heidelberg 1, Germany
\and Abastumani Astrophysical Observatory, Georgia
\and Astrophysikalisches Institute Potsdam, An der Sternwarte 16, D-14482
Potsdam, Germany 
\and Dipartimento di Fisica, Universit\`a La Sapienza, Roma,
P.le A.\ Moro 2, I-00185 Roma, Italy
\and Tuorla Observatory, FIN-21500 Piikki\"o, Finland 
\and  Dipartimento di Fisica Generale, Universit\`a di Torino, Via P.\ Giuria
1, I-10125 Torino, Italy 
\and Mets\"ahovi Radio Observatory, Mets\"ahovintie 114, FIN-02540
Kylm\"al\"a, Finland 
\and Cattedra di Astrofisica, Universit\`a di Perugia, Via B.\ Bonfigli,
I-06126 Perugia, Italy 
\and Foggy Bottom Observatory, Colgate University, Hamilton, NY, USA
\and Max-Planck-Institut f\"ur Astronomie, K\"onigstuhl 17, D-69117
Heidelberg, Germany 
\and Mount Stromlo and Siding Spring Observatories, Canberra,
Australia 
\and Beijing Astronomical Observatory, Beijing, China
\and Teide Observatory, Tenerife, Spain
\and  Dept.\ of Chemistry, Physics, \&
Astronomy, Francis Marion University, P.O.\ Box 100547, Florence, SC
29501-0547, USA   
\and Vainu Bappu Observatory, Indian Institute of Astrophysics,
Dept.\ of Science \& Technology, Kavalur, India
\and Dept.\ of Physics and Astronomy, University of Victoria, British Columbia,
Canada 
\and Dept.\ of Physics, University of Colorado at Denver, P.O.\ Box 173364,
Denver, CO  80217-3364, USA }

\date{Received; Accepted;}

\titlerunning{Optical and radio variability of AO 0235+16}

\authorrunning{C.M.\ Raiteri et al.}

\abstract{ 
The BL Lacertae object \object{AO 0235+16} is well known for its extreme
optical and radio variability. New optical and radio data have been collected
in the last four years by a wide international collaboration, which confirm
the intense activity of this source: on the long term, overall variations
of $5\rm\,mag$ in the $R$ band and up to a factor $18$ in the radio
fluxes were detected, while short-term variability up to $0.5\rm\,mag$ in a few
hours and $1.3\rm\,mag$ in one day was observed in the optical band.  The optical
data also include the results of the  Whole Earth Blazar Telescope (WEBT)
first-light campaign organized in November 1997, involving a dozen optical
observatories. The optical spectrum is observed to basically steepen when the
source gets fainter.
We have investigated the existence of typical variability time
scales and of possible correlations between the optical and radio emissions by
means of visual inspection and Discrete Correlation Function (DCF)
analysis. 
On the long term, the autocorrelation function of the optical data shows a
double-peaked maximum at $4100$--$4200$ days ($11.2$--$11.5$ years), while
a double-peaked maximum at $3900$--$4200$ days ($10.7$--$11.5$ years) is visible in the
radio autocorrelation functions. The existence of this similar
characteristic time scale of variability in the two bands is by itself an
indication of optical-radio correlation. A further analysis by means of
Discrete Fourier Transform (DFT) technique and folded light curves reveals that
the major radio outbursts repeat quasi-regularly with a periodicity of $\sim
5.7$ years, i.e.\ half the above time scale. This period is also in agreement
with the occurrence of some of the major optical outbursts, but not all of
them. Visual inspection and DCF analysis of the
optical and radio light curves then reveal that in some cases optical
outbursts seem to be simultaneous with radio ones, but in other cases they 
lead the radio events. Moreover, a deep inspection of the radio light curves
suggests that in at least two occasions (the 1992--1993 and 1998 outbursts)
flux variations at the higher frequencies  may have led those at the lower
ones.    
} 

\maketitle
\section{Introduction}

BL Lacertae objects, together with quasars with flat radio spectrum, belong to
the class of active galactic nuclei called blazars. These are objects
characterized by extreme emission variability at all wavelengths, noticeable
emission at the $\gamma$-ray energies, important polarization, superluminal
motion, brightness temperatures by far exceeding the Compton limit (see e.g.\
Urry \cite{urr99} for a review). 

The general scenario for the blazar emissivity has long been set: it foresees
a plasma jet coming out from a supermassive, rotating black hole surrounded by
an accretion disk. In the jet, which is closely aligned with our line of
sight, relativistic electrons produce soft photons (from radio to
the UV--X-rays) through synchrotron emission, and high-energy photons (up to
the TeV energies) via inverse Compton scattering. However, theoretical models
differ in the jet structure and physics, and in the origin of the seed photons
that are energized to the $\gamma$-ray energies. Constraints on the
theoretical models can come from dense optical and radio monitoring, and
simultaneous observations throughout all the electromagnetic spectrum, from
the radio to the $\gamma$-ray band. Indeed, multiwavelength variability studies
(see  Ulrich et al.\ \cite{ulr97} for a review) can give information
on the compactness of the emitting regions, and verify the
existence of correlations and time delays among the emissions in different
bands. This in turn can
shed light on the location of the emitting regions in the jet and on the
nature of the seed photons that are comptonized. International collaborations
among optical astronomers such as the OJ-94 Project and the Whole Earth Blazar
Telescope (WEBT) were born in the last years to make the observational effort
more efficient.

In this paper we present the results of a wide collaboration aimed to
study the optical and radio variability of the BL Lac object AO 0235+16.
Partial and very preliminary results were presented in Villata et al.\
(\cite{vil99a}), where the light curve in the $R$ band from January 1993 to
January 1999 was shown and radio-optical correlations during that period
analyzed by means of the Discrete Correlation Function (DCF).

The source AO 0235+16 is a well known blazar at $z=0.94$ that exhibits
spectacular emission variations on both short ($\sim$ day) and long (months,
years) time scales. 

It was classified as a BL Lac object by Spinrad \& Smith (\cite{spi75}), who
observed a $\sim 2$ mag variation. Noticeable optical outbursts were
described by Rieke et al.\ (\cite{rie76}), Pica et al.\ (\cite{pic80}), Webb
et al.\ (\cite{web88}), Webb \& Smith (\cite{web89}), Webb et al.\
(\cite{web00}). 
From the data reported in the literature, an overall brightness
variation of more than $5\rm\,mag$ can be seen, which makes AO 0235+16 one of
the most interesting sources for variability studies.

As for the short-term optical behaviour, 
intraday variability was observed in a number of occasions (Heidt \& Wagner
\cite{hei96}; Noble \& Miller \cite{nob96}; Romero
et al.\ \cite{rom00}). In particular, changes of $0.5\rm\,mag$ were measured by
Romero et al.\ (\cite{rom00}) within a single night, and variations of more
than $1\rm\,mag$ in about $24$ hours.

Many optical data were taken in the period 1993--1996 by the astronomers
participating in the international collaboration called OJ-94 Project,
who observed AO 0235+16 as a ``complementary" object, in addition to \object{OJ 287}
and \object{3C 66A}. The results of their monitoring were published in Takalo et al.\
(\cite{tak98}).
Some of the groups belonging to the OJ-94 Project have then
continued to observe AO 0235+16.

A huge monitoring effort has been
pursued in the radio band by the University of Michigan Radio Observatory
(UMRAO), in the USA, and by the Mets\"ahovi Radio Observatory, in Finland. 
Indeed, this object is of extreme interest also in the radio band.
Long-term radio light curves at frequencies above $1\rm\,GHz$ have shown that
it exhibits frequent, relatively well-resolved, high-amplitude events
(O'Dell et al.\ \cite{ode88}; Clements et al.\ \cite{cle95}). This makes
AO 0235+16 particularly suitable for studies of correlated activity.
Moreover, several recent papers have argued that unusually high Doppler
factors/large Lorentz factors may be present: Jorstad et al.\ (\cite{jor01})
infer a Lorentz factor greater than $45$ based on $43\rm\,GHz$ proper motions, using
VLBA observations during 1995 and 1996; a Doppler factor for the bulk flow of
order $100$ was inferred from radio short-term variability in an earlier paper 
by Kraus et al.\ (\cite{kra99}). Doppler
factors larger than $80$ (and Lorentz factors $\Gamma > 60$) were found
by Fujisawa et al.\ (\cite{fuj99}) by applying the inverse-Compton effect and
equipartition models to VLBI observations at $22\rm\,GHz$. Recent VSOP
observations of AO 0235+16 by Frey et al.\ (\cite{fre00}) have led to an
estimate of $T_{\rm b} > 5.8 \times 10^{13} \rm \, K$ for the rest-frame brightness
temperature of the core, which is the highest value measured with VSOP to
date and implies a Doppler factor of $\sim 100$.

Optical and radio data on AO 0235+16 taken in the last four years
(1996--2000) by a wide international collaboration are presented in Sect.\ 2,
and inserted in the source long-term light curves, dating back from the mid
Seventies. 
The results of the Whole Earth Blazar Telescope (WEBT) first-light campaign
on AO 0235+16 in autumn 1997 are also included.
Optical spectra are derived and spectral changes discussed in the same section.
Radio light curves are shown in Sect.\
3, where the behaviour of the radio flux is analyzed side by side with the
optical one.
Statistical analysis is presented in Sect.\ 4: autocorrelation function and
Discrete Fourier Transform (DFT) methods are applied to search for
characteristic time scales of variability in both the radio and optical
domains; DCF analysis is then performed to
check the existence of radio-optical correlations.
A final discussion is contained in Sect.\ 5.

\section{Optical behaviour of AO 0235+16}

\begin{figure*} 
\epsfxsize=15cm \epsfbox{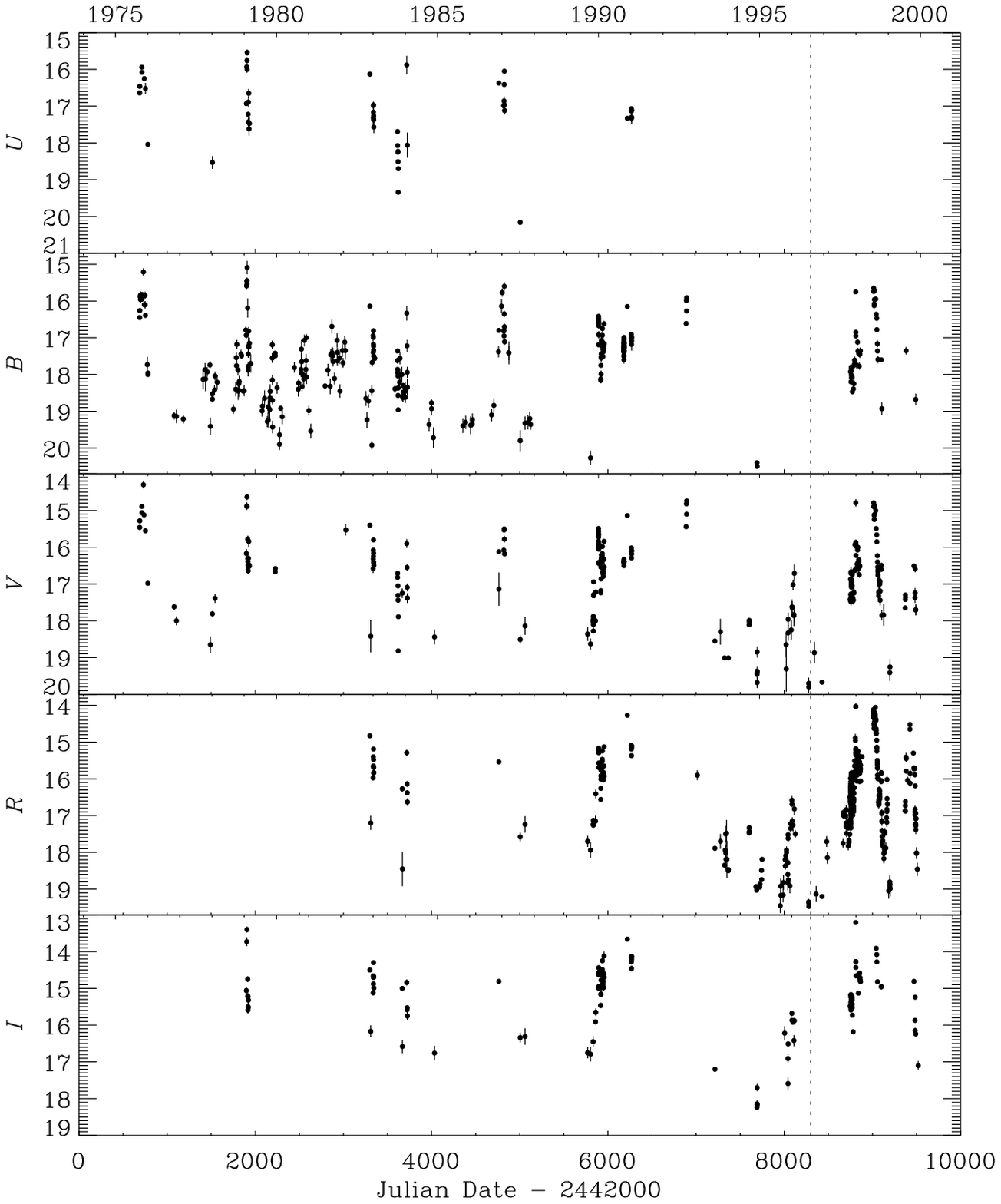} 
\caption[]{Long-term monitoring light curves of AO 0235+16 in $UBVRI$; data are
from Kinman \& Rieke (\cite{kin75}),
Rieke et al.\ (\cite{rie76}), O'Dell et al.\ (\cite{ode78a}, \cite{ode78b}),
Pica et al.\ (\cite{pic80}), Impey et al.\ (\cite{imp82}), Barbieri et al.\
(\cite{bar82}), Moles et al.\ (\cite{mol85}), Smith et al.\ (\cite{smi87}),
Sillanp\"a\"a et al.\ (\cite{sil88a}), Webb \& Smith (\cite{web89}), Mead
et al.\ (\cite{mea90}), Sitko \& Sitko (\cite{sit91}), Sillanp\"a\"a et
al.\ (\cite{sil91}), Takalo et al.\ (\cite{tak92}), Xie et al.\
(\cite{xie92}), Rabbette et al.\ (\cite{rab96}), Webb et al.\
(\cite{web97}), Takalo et al.\ (\cite{tak98}), Xie et al.\ (\cite{xie99}), and
Ghosh et al.\ (\cite{gho00}); data plotted after the vertical line are from
the present work}                                                            
\label{ubvri}
\end{figure*}

\begin{figure*} 
\epsfxsize=15cm \epsfbox{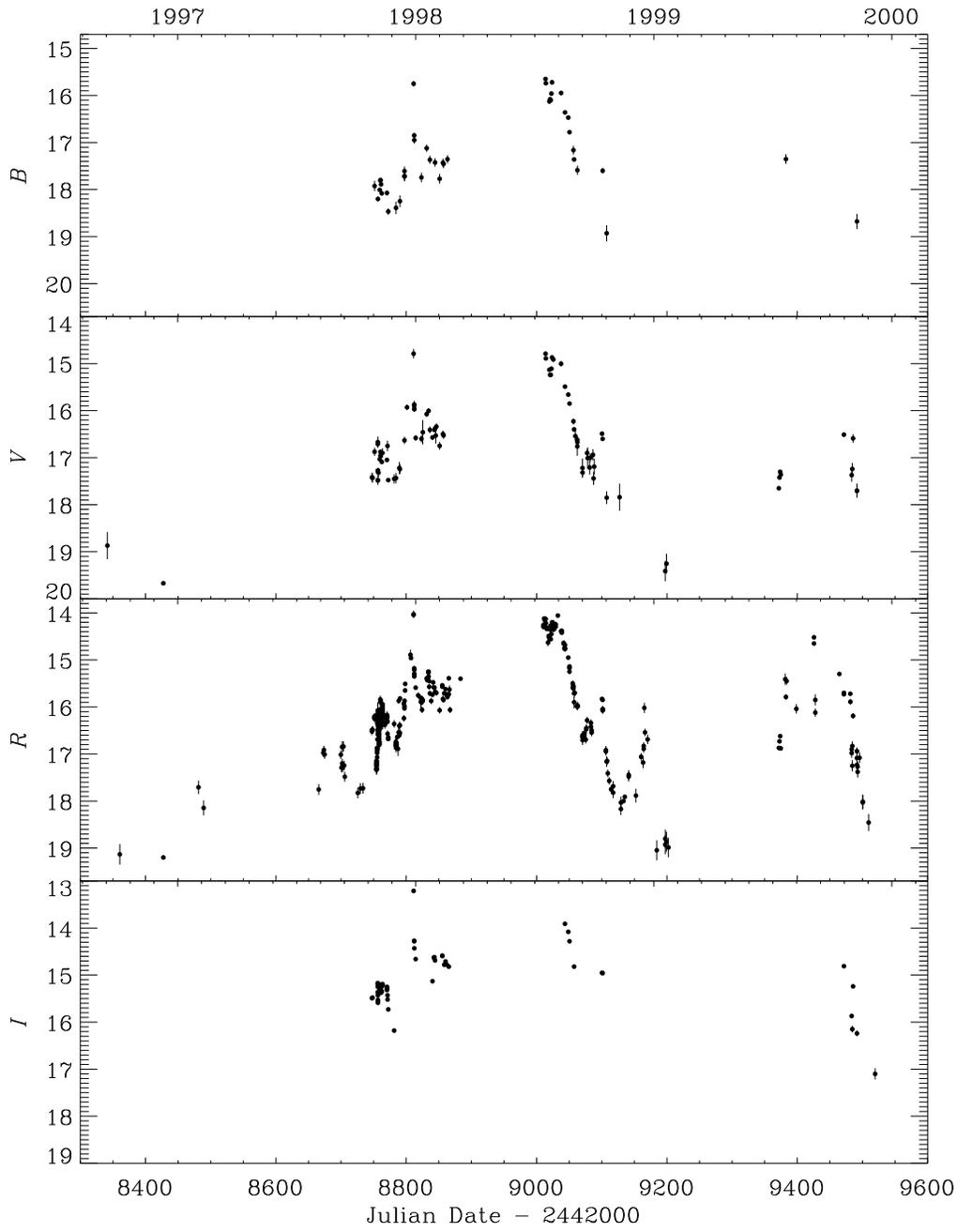} 
\caption[]{Light curves of AO 0235+16 in $BVRI$ bands in the years 1996--2000;
all data are from the present work} 
\label{bvri}     
\end{figure*}

\begin{table*} 
\caption{List of participating observatories by longitude} 
\begin{tabular}{lllllll} 
\hline 
Observatory & Nation  & Long.\ (deg) & Lat.\ (deg) & Tel.\
diam.\ (m)& Filters & Symbol$^a$\\  
\hline  
MSSSO   & Australia & $+149.066$ & $-31.277$ & $1.0$ &$R$& filled triangles\\   
Xinglong    & China  & $+117.575$ & $+40.393$ & $0.6$ &$R$& filled squares  \\
Vainu Bappu & India & $+78.83$ & $+12.57$ & $2.0$ &$R$& asterisks       \\  
Abastumani  & Georgia & $+42.80$ & $+41.80$ & $0.7$ &$R$ & open circles  \\
Tuorla      & Finland & $+22.17$ & $+60.27$ & $1.03$ & $V$ &   \\  
Roma        & Italy  & $+12.70$ & $+41.80$ & $0.5$, $0.7$ &$BVRI$ & open triangles\\  
Perugia     & Italy  & $+12.372$ & $+43.112$ & $0.4$ &$VRI$  & open diamonds\\  
Torino      & Italy  & $+7.775$ & $+45.038$ & $1.05$ &$BVRI$ & open squares\\      
Calar Alto  & Spain  & $-2.546$ & $+37.224$ & $2.2$ &$R$    & filled circles\\    
Teide       & Spain  & $-16.5$ & $+28.16$ & $0.82$ &$BVRI$    & crosses \\ 
NOT         & Spain  & $-17.88$ & $+28.75$ & $2.56$ &$BVRI$ & asterisks \\
Foggy Bottom& USA    & $-74.25$ & $+43.24$ & $0.4$ &$R$    & filled stars   \\  
Climenhaga  & Canada & $-123.22$ & $+48.25$ & $0.5$ &$R$    & filled diamonds \\ 
\hline  
\end{tabular}

\vspace{0.2cm}\noindent
$^a$ The symbols are used to distinguish the $R$ band data plotted in Figs.\
3--7   
\end{table*}

The long-term monitoring optical light curves of AO 0235+16 in the Johnson's
$UBV$ and Cousins' $RI$ bands are shown in Fig.\ 1.
As generally happens, the best sampled band in the
past is the $B$ one, while the use of CCD cameras has led people to
observe chiefly in the $R$ band. 
The data plotted before the vertical
line at ${\rm JD}=2450300$ were derived from the literature (see references in
the caption); those after are new observations
presented in this paper. Thirteen optical observatories all over the world
were involved in this project; their names, nationality, position
(longitude and latitude), diameter of the telescope and filters used are
indicated in Table 1. Seven of them (MSSSO, Xinglong, Vainu Bappu, Calar Alto,
Teide, Foggy Bottom, Climenhanga) participated only during the WEBT campaign in
November 1997 (see Sect.\ 2.1) or around that period, while the others
provided the long-term optical monitoring.
All observers took frames by means of CCD cameras and performed data
reduction using standard procedures; the source magnitude
was obtained by differential photometry, adopting the reference stars
calibrated by Smith et al.\ (\cite{smi85}) and Fiorucci et al.\
(\cite{fio98}). 

Although the intense source variability makes a comparison among the
different datasets difficult to perform, a general agreement was found
where the data overlap in time.

Many large-amplitude outbursts are visible from Fig.\ 1; in the $B$ band the
minimum and maximum magnitudes observed were $15.09\pm 0.18$ (${\rm JD}=2443907$)
and $20.5\pm 0.05$ (${\rm JD}=2449692$), respectively. In the $R$ band
the range of magnitudes spanned is from $14.03 \pm 0.08$ (${\rm JD}=2450811$) to
$19.47 \pm 0.06$ (${\rm JD}=2450278$).

An enlargement of the $BVRI$ light curves in the period 1996--2000 is shown in
Fig.\ 2; an inspection of the $R$-band curve, the best sampled one, reveals
three major peaks, one for each of the last three observing seasons.

Details of the 1997--1998 season are visible in Fig.\ 3, where different
symbols refer to different observatories (see caption to the figure), with
the exception of dark dots, which indicate data collected during the WEBT
campaign.

The peak at ${\rm JD}=2450811$ was observed by both the Perugia and the Roma
Observatories, and corresponding peaks were seen also in the $BVI$ bands (see
Figs.\ 1 and 2). What is quite impressive is the extreme sharpness of the
flare: a brightness increase of $1\rm\,mag$ in $4$ days (inside a $2.8\rm\,mag$ rise in $27$
days) was followed by a $1.3\rm\,mag$ decrease in only one day, leading to a
$2.0\rm\,mag$ fall in $13$ days.

A smoother behaviour characterizes the 1998--1999 observational season (see
Fig.\ 4): the source was found to be very bright at the beginning of the
season, skimming $R=14$, and then a continuous decrease, interrupted by a
couple of flares, led to $R\sim 19$, with a jump of $5\rm\,mag$. 

In the last observing season the temporal coverage was definitely worse than in
the previous two seasons (see Fig.\ 2); however, a quite interesting feature is
the fall of $1.6\rm\,mag$ in $48$ hours occurred in the $R$ band at the beginning of
September 1999 (${\rm JD}=2451425.657$--$2451427.657$).

\begin{figure} 
\hspace{-0.3cm}\epsfxsize=10cm \epsfbox{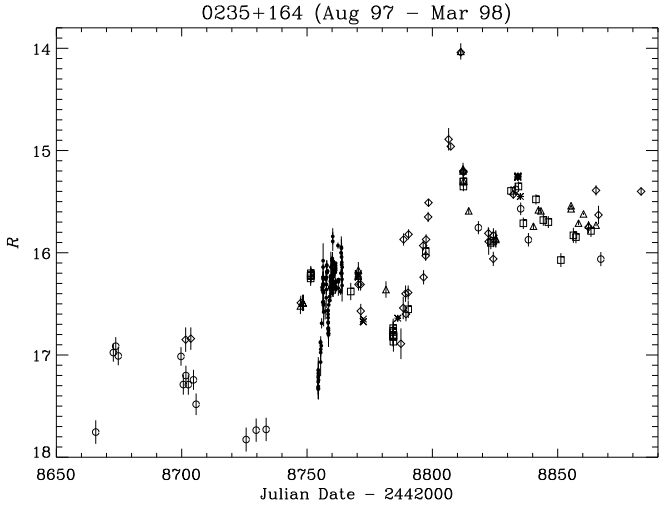} 
\caption[]{Light curve of AO 0235+16 in the Cousins' $R$ band from 
August 1997 to March 1998; data are from Vainu Bappu (asterisks), Abastumani 
(open circles), Roma
(triangles), Perugia (diamonds), Torino (squares), and Teide (crosses);
data from the WEBT campaign occurred from November 1 to 11, 1997
(${\rm JD}=2450754$--$2450764$) are plotted as dark dots} 
\label{r1}     
\end{figure} 

\begin{figure} 
\hspace{-0.3cm}\epsfxsize=10cm \epsfbox{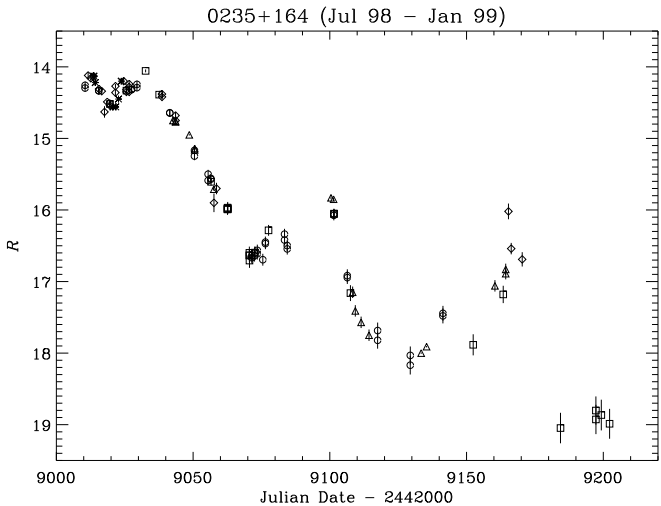} 
\caption[]{Light curve of AO 0235+16 in the Cousins' $R$ band from 
July 1998 to January 1999; data are from Abastumani (circles), Roma
(triangles), Perugia (diamonds), Torino (squares), Teide (crosses), and
NOT (asterisks)} 
\label{r2}     
\end{figure}

\subsection{The November 1997 WEBT campaign}

The WEBT is an international collaboration among 
astronomers from all the world with the aim of organizing optical campaigns 
on blazars of specific interest (Mattox \cite{mat99a}, \cite{mat99b};
Villata et al.\ \cite{vil00}). These campaigns, lasting from a few  days to
several weeks, are generally undertaken in concert with observations at other
wavelengths, and can be triggered by the discovery of a flaring state of an
object, usually in the optical band. 

This was indeed the case for the first-light WEBT
campaign: the  November 1997 observations of AO 0235+16 were started
after the detection of a considerable brightness increase (about $1\rm\,mag$ in a
week) at the end of October 1997 (Webb et al.\ \cite{web97}).    
An EGRET target of opportunity observation occurred between November 3 and 11,
1997, but only an upper limit of $16 \times 10^{-8} \, \rm photons \, cm^{-2}
\, s^{-1}$ above $100\rm\,MeV$ could be inferred (Hartman, private communication).  
Observations by RXTE did not find a high X-ray flux (Webb et al.\
\cite{web00}). 

The light curve obtained in the first $11$ days of November 1997 is plotted in
Fig.\ 5. Data from each observatory are indicated  with a different symbol,
according to Table 1 (Column 7). Weather was rather bad in those days in most
of western Europe and North America, so that the temporal density of the
curve is far from ideal, and far even from the density reached in more recent
WEBT campaigns (see Villata et al.\ \cite{vil00}, \cite{vil01}); however, the
common observational effort of many observatories around the world led to a
satisfactory time coverage.

In order to avoid spurious variations due to possible systematic effects
among data from different observatories, in the following only flux changes
observed by the same telescope will be considered. A brightness increase of
$1.25\rm\,mag$ in two days was detected at the beginning of the campaign and
noticeable flux oscillations were observed all the time. Variations of
$0.2$--$0.4\rm\,mag$ in $5$--$7$ hours were found in a number of cases; very impressive is
the dimming of half a magnitude in about $5$ hours  detected in the night
between November 3 and 4. Similar very fast variations have recently been
observed by Romero et al.\ (\cite{rom00}).

\begin{figure*} 
\epsfxsize=15cm \epsfbox{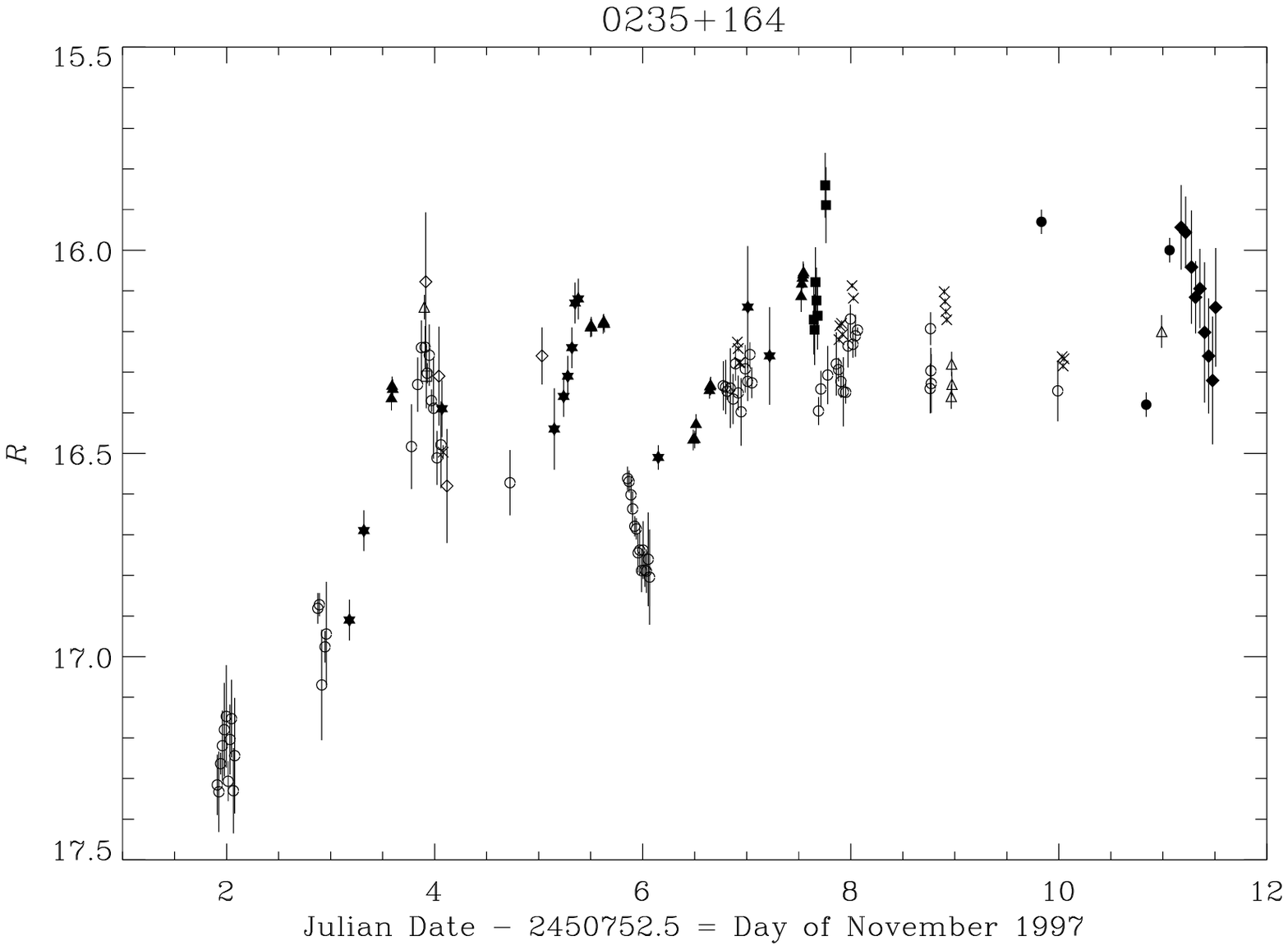} 
\caption[]{Light curve of AO 0235+16 in the Cousins' $R$ band during the
first-light WEBT campaign of November 1997; for the explanation of
symbols see Table 1}  
\label{cltot}
\end{figure*}

\subsection{Spectral behaviour in the optical band}

\begin{figure} 
\epsfxsize=9cm \epsfbox{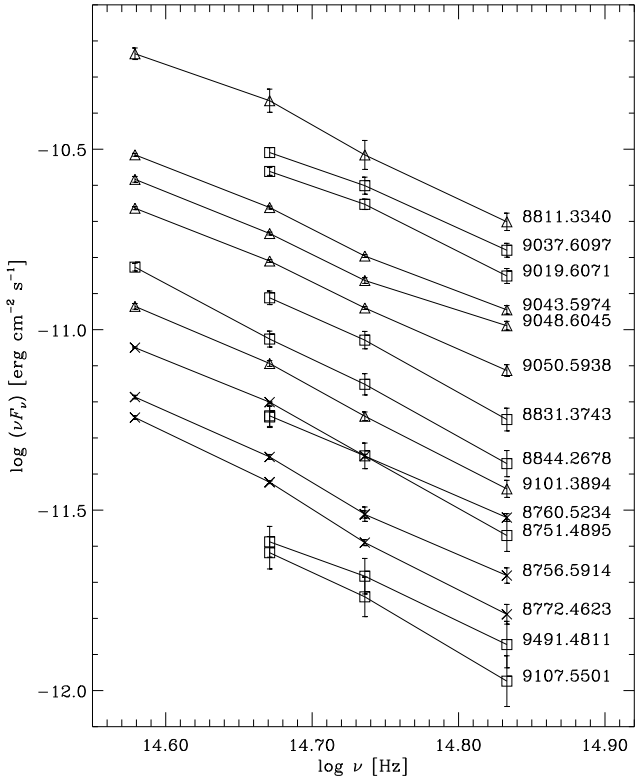} 
\caption[]{A selection of optical spectra from Roma (triangles), Torino
(squares), and Teide (crosses); Julian Dates ($-2442000$) are indicated on the
right; straight lines are drawn only to guide the eye through points of the
same spectrum; no vertical shift is applied}    
\label{spettrisel} 
\end{figure}

One item that is interesting to investigate when studying the optical
variability of blazars is whether the flux variations are accompanied by 
spectral changes. 
In the new data presented in this paper we identified 47 optical spectra
collecting data in at least three filters taken by the same telescope,
with the requirement that the delay from the first frame to the last one is
not greater than $50$ minutes.  
We discarded 7 spectra which contained large
errors and analyzed the remaining 40. 
A selection of these ``well behaved" spectra is
presented in Fig. \ref{spettrisel} in the $\log \nu$ - $\log (\nu
F_{\nu})$ plane, using different symbols for data taken at different
telescopes. When taking into account that some discrepancies in the spectral
shape can arise from not completely equal photometric systems used in the
various observatories, no significant variations can be recognized when the
flux changes. However, a deeper inspection reveals that the spread
corresponding to the $B$ band is larger than that relative to the $I$ one.
This would suggest that the flux variations are of larger amplitude at the
higher frequencies, as already observed in other blazars. 
 
\begin{figure} 
\hspace{-0.3cm}\epsfxsize=9cm \epsfbox{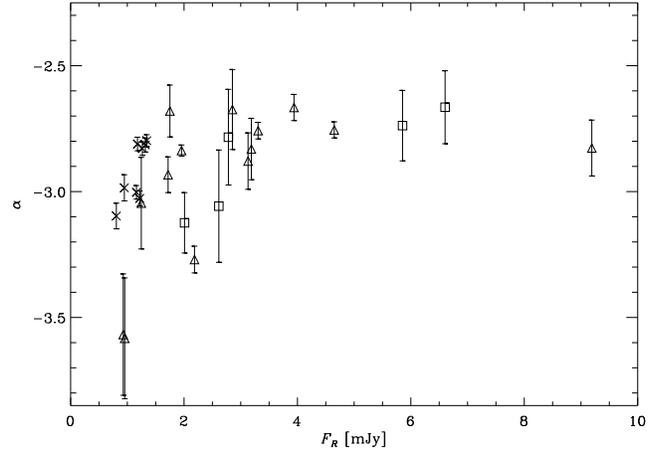} 
\caption[]{The spectral index $\alpha$ as a function of the $R$ flux for the
40 optical spectra from Roma (triangles), Torino (squares), and Teide
(crosses)}    
\label{alfa}
\end{figure}

All 40 ``well behaved" optical spectra were then fitted by a classical power
law: $F_{\nu} \propto \nu^{\alpha}$ with a $\chi^2$ minimizing procedure. The
results are shown in Fig.\ \ref{alfa}, where the spectral index $\alpha$ is
plotted as a function of the flux in the $R$ band and $\alpha$ values with
errors greater than $0.25$ have been eliminated. One can notice that for high
flux levels $\alpha$ presents an almost constant value of $-2.8$--$-2.7$,
while it tends to decrease with
decreasing flux, although there is a not negligible spread. 
This means that the spectrum basically steepens when the source gets fainter, a
behaviour that is common to blazars.

\section{Radio and optical fluxes: visual correlation analysis}
 
\begin{figure*}  
\epsfxsize=15cm \epsfbox{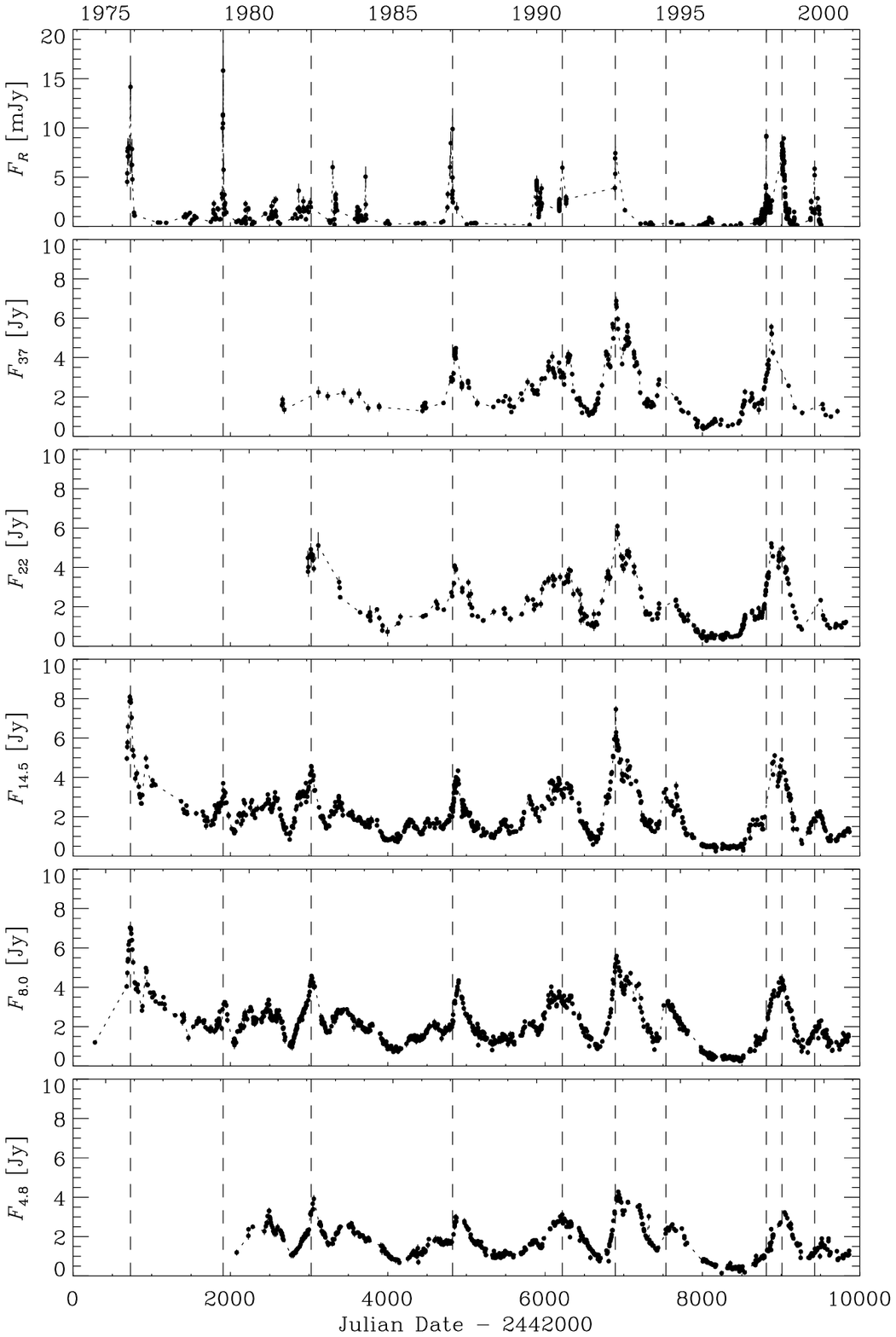} 
\vspace{1.0cm}
\caption[]{Optical (mJy) and radio (Jy) light curves of AO 0235+16;  radio
data at $22$ and $37\rm\,GHz$ are from the Mets\"ahovi Radio Observatory,  those at
$14.5$, $8.0$, and $4.8\rm\,GHz$ are from UMRAO}
\label{radop}     
\end{figure*}

Radio light curves of AO 0235+16 are shown in Fig.\ \ref{radop}: data at $22$ and
$37\rm\,GHz$ are from the Mets\"ahovi Radio Observatory, those at $4.8$, $8.0$, and 
$14.5\rm\,GHz$ are from the University of Michigan Radio Astronomy Observatory (UMRAO). 
The observations at the Mets\"ahovi Radio Observatory were made with the $13.7\rm\,m$
antenna using standard ON/ON techniques with DR 21 used as a calibration
source. Details on the observing procedure and data reduction can be found in
Ter\"asranta et al.\ (\cite{ter92}, \cite{ter98}).
The data from UMRAO were taken with the $26\rm\,m$ paraboloid of the University
of Michigan. A description of the observing and data reduction procedures is
included in Aller et al.\ (\cite{all85}, \cite{all99}).

As in the optical band, also in the radio AO 0235+16 presents intense activity
at all wavelengths, with pronounced outbursts lasting from several months to a
few years. The overall flux variations (maximum value over minimum one)
detected in the various bands are: $18$, $22$, $32$, $28$, and $31$ at $37$, $22$, $14.5$,
$8.0$, and $4.8\rm\,GHz$, respectively. In particular, in the last four years (${\rm
JD} > 2450300$) variations up to a factor $18$ were observed.

In Fig.\ \ref{radop} radio fluxes (in Jy) are compared to the optical ones
(in mJy), obtained in the following way: all magnitudes in
the $B$ band before ${\rm JD}=2449000$ have been transformed into $R$ ones
adopting the mean colour index $<B-R>\ =1.65 \pm 0.16$ (which was
derived by considering all the $B$-$R$ pairs from a same observatory
separated by no more than half an hour); 
these data plus the real $R$ magnitudes after
${\rm JD}=2449000$ have been converted into fluxes by adopting Rieke \& Lebovski
(\cite{rie85}) and the law by Cardelli et al.\ (\cite{car89}), and using a
Galactic extinction $A_B=0.341 \, \rm mag$ (from NED).

The existence of radio-optical correlations for AO 0235+16 was investigated in
a number of previous works; evidence for a simultaneous radio and optical
variability was found in correspondence to the optical flares of 1975 and 1979
(MacLeod et al.\ \cite{mac76}; Ledden et al.\ \cite{led76}; Rieke et al.\
\cite{rie76}; Balonek \& Dent \cite{bal80}) and, more recently, to that which
occurred in 1997 (Webb et al.\ \cite{web00}).

Clements et al.\ (\cite{cle95}) analyzed optical data taken in the period
1977--1991 and radio data at $8.0\rm\,GHz$ from UMRAO with the Discrete
Correlation Function (DCF), and found that: ``Overall, radio events lag optical
events with lag times varying from $0$ to $2$ months.".
Takalo et al.\
(\cite{tak98}) visually compared optical data from 1980 to 1996 with combined
$22$ and $37\rm\,GHz$ data from the Mets\"ahovi Radio Observatory and noticed that
some of the optical spikes appear to be coincident with radio flares, while
others have no counterparts. Moreover, the general trend looked very similar
in both frequency regimes, suggesting some kind of correlation.

A visual inspection of Fig.\ \ref{radop} shows that the big optical outburst
of 1975 has a big radio counterpart at $14.5$ and $8.0\rm\,GHz$, while in 1979 a
noticeable optical peak corresponds to a modest radio peak. 
Prominent radio outbursts at $22$, $14.5$, $8.0$, and $4.8\rm\,GHz$ were observed in 1982,
towards the end of the optical season, so that a possible optical peak might 
have been missed. A strong brightness increase was detected in 1987 in both
optical and radio bands, but in this case the double-peaked optical
flare seems to preceed the radio ones.  The behaviour of the
long radio outburst of 1990--1991 appears more complex; during the
radio outburst a sharper optical flare was detected but not followed in
details. The radio outburst in 1992--1993 was double-peaked; just before the
first radio peak, an optical flare was detected; no other optical data were
taken at the time when the radio fluxes reached their maxima. One
interesting feature, however, is that if one looks at the better-sampled UMRAO
data, the first radio peak seems delayed  when proceeding from the higher to
the lower radio frequencies. Indeed, the maximum value
was reached on October 4, 1992 at $14.5\rm\,GHz$, on  October 13 at $8.0\rm\,GHz$, and on
November 3 at $4.8\rm\,GHz$.
A radio flux increase in 1994 was practically not
followed in the optical band. 

From these considerations it is clear that the main difficulty in performing a
meaningful study on possible radio-optical correlations is the paucity of
optical data. 

\begin{figure*}  
\epsfxsize=15cm \epsfbox{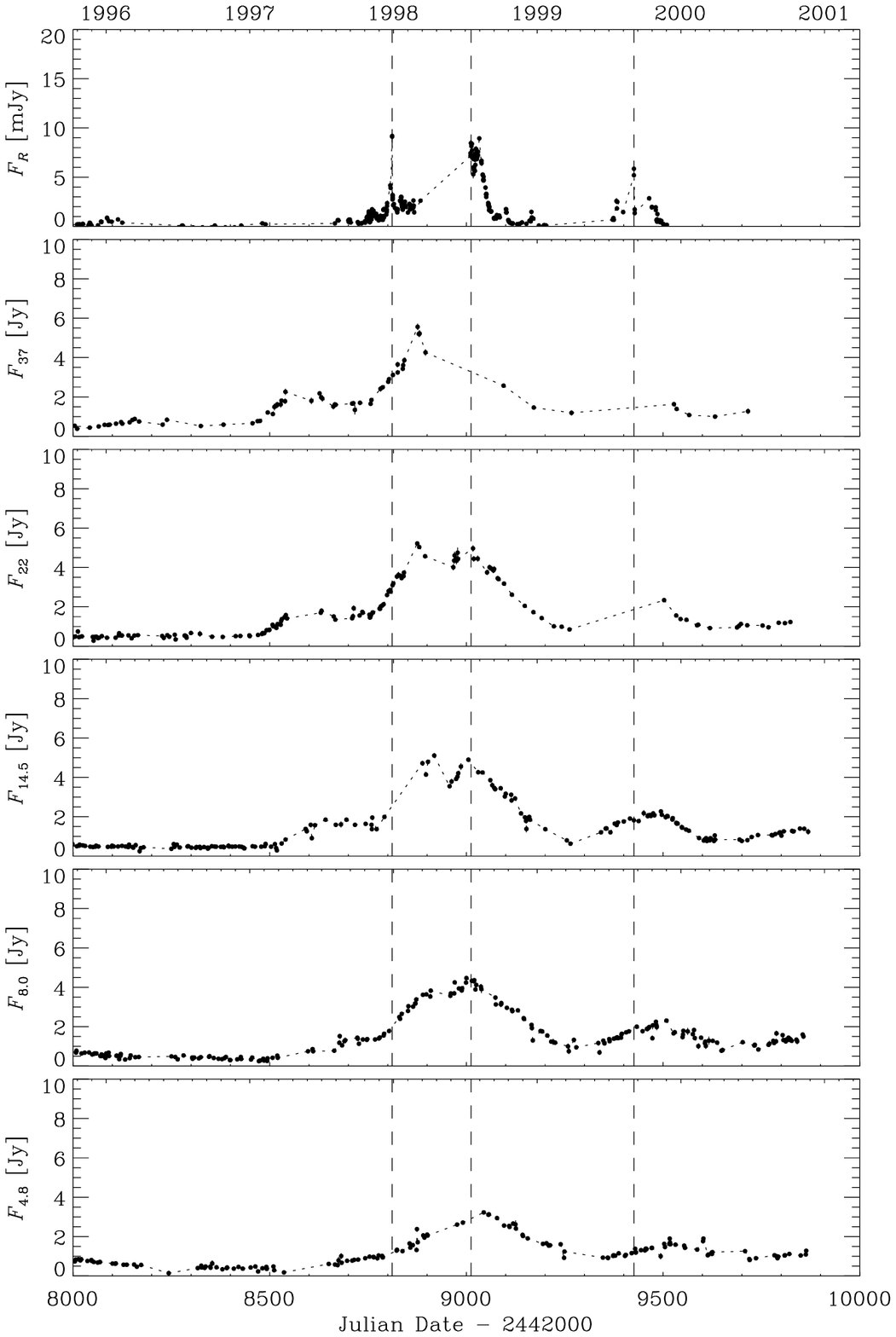}
\vspace{1.0cm} 
\caption[]{Optical (mJy) and radio (Jy) light curves of AO 0235+16 in the
last years;  radio data at $22$ and $37\rm\,GHz$ are from the Mets\"ahovi
Radio Observatory,  those at $14.5$, $8.0$, and $4.8\rm\,GHz$ are from UMRAO}
\label{radopz8}     
\end{figure*}

The situation has been noticeably improved in the last years, because of
the intense observational effort of the monitoring groups involved in the
present work. Indeed, the big outbursts occurred at the end of 1997 and in 1998
were accurately followed in all the radio bands and in the optical one (Fig.\
\ref{radopz8}).

A preliminary discussion on the radio-optical correlations
in this period was presented by Villata et al.\ (\cite{vil99a}).
 
On December 28, 1997 (${\rm JD}=2450811.33$), the $R$-band flux reached
$9.19 \pm 0.68 \, \rm mJy$ ($R=14.03$); four days before it was  $3.90 \pm 0.11
\, \rm mJy$ and $24$ hours after the peak the flux had dropped to $2.85 \pm 0.13
\, \rm mJy$. This very sharp peak was observed by two different groups (Perugia
and Roma) and also in different bands (see Figs.\ \ref{ubvri} and \ref{bvri}).
At the time of this optical flare, the radio fluxes were in a rising phase:
they reached their maximum values later,  likely first the shorter (observed by
Mets\"ahovi) and then the longer (UMRAO) wavelengths. The peaks at $37$ and $22\rm\,GHz$ 
were detected on March 3 and 2, 1998, respectively,
although a gap in the radio datasets does not allow to establish whether the
peak of the outburst had already occurred. The maximum value of the $14.5\rm\,GHz$
flux was seen on April 14, 1998 (${\rm JD}=2450918$), and that at $8.0\rm\,GHz$ on
April 5, 1998, but in this case the peak probably occurred later. 
This peak-delay effect has already been quoted above for the 1992--1993
outburst. The radio
outburst at $22$ and $14.5\rm\,GHz$ is clearly double-peaked; this feature is
recognizable also in the $8.0\rm\,GHz$ data.  It is noticeable that the second radio
peak could reasonably be contemporaneous at $22$, $14.5$, and $8.0\rm\,GHz$, the best
sampled bands, and corresponds to a second optical flare detected at the
beginning of the observing season, in July--August 1998. The solar
conjunction period prevented to follow the rising phase of the optical
outburst, so that the possibility that the optical peak also in this case
preceded the radio ones remains open. A second brighter optical peak detected
on August 7, 1998 (${\rm JD}=2451032.63$, $R$ flux of $8.94 \pm 0.16 \, \rm mJy$)
occurred when the radio flux was in a decreasing stage.  Another important,
sharp optical flare was finally detected on September 4, 1999
(${\rm JD}=2451425.66$), which again has a radio counterpart, whose peak shows
several weeks of delay.

The above discussion demonstrates that, notwithstanding the great observational
effort of the last years, we are still far from having the sufficient sampling
to derive firm conclusions on the radio-optical correlations. We can only
notice that, in general, when the observational coverage is sufficiently good,
a long time scale radio flux increase corresponds to a short time scale
optical brightness increase, whose peak may precede the radio one.
Moreover, there are at least two cases (the 1992--1993 and 1998
double-peaked outbursts) where a progressive time delay in reaching the maximum
value is observed when passing from the higher to the lower frequency radio
fluxes.

\section{Statistical analysis}

In this section we apply the Discrete Correlation Function
(DCF) analysis to the data shown in Fig.\ \ref{radop} in order to investigate
the existence of characteristic time scales of variability and of optical-radio
correlations.

The DCF is a method specifically
designed for unevenly sampled datasets (Edelson \& Krolik \cite{ede88};
Hufnagel \& Bregman \cite{huf92}), which also allows an estimate of the
accuracy of its results.

Given two datasets $a_i$ and $b_j$, one has first to combine all pairs,
calculating the unbinned discrete correlations:
$${\rm UDCF}_{ij}={{(a_i- \overline a)(b_j- \overline b)} \over {\sigma_a
\sigma_b}},$$
where $\overline a$, $\overline b$ are the average values of the two datasets,
and $\sigma_a$, $\sigma_b$ their standard deviations. The DCF is obtained by
binning the ${\rm UDCF}_{ij}$ in time for each time lag $\tau$:
$${\rm DCF} (\tau)= {1 \over M} \sum {\rm UDCF}_{ij}(\tau),$$
where $M$ is the number of pairs $a_i, b_j$ whose time lag $\Delta_{ij}=t_j -
t_i$ is inside the $\tau$ bin. Spurious correlations can be found, of the
order of $\pm M^{-1/2}$.
The standard error for each bin is:
$$\sigma_{\rm DCF} (\tau) = {1 \over {M-1}} \left\{ \sum \left[ {\rm
UDCF}_{ij} - {\rm DCF}(\tau) \right] ^2 \right\}  ^{1/2}.$$

A positive peak of the DCF means correlation, which is stronger as the value
of the peak approaches and exceeds one. A negative peak implies
anticorrelation. Moreover, the width of the peak must be comparable to those
of the autocorrelation functions, obtained by applying the DCF to each
dataset coupled with itself.  

A preliminary binning of data in time before calculating the DCF usually leads
to better results, smoothing out flickering. The size of this binning is
crucial especially in the optical, where short-term variations are frequent, 
since it can remove important information. 
Furthermore, an increase of the data binning interval implies an increase of
the spurious correlations, while an increase of the DCF bin size has the
opposite effect. Also the choice of the DCF binning is a delicate point,
determining the balance between resolution and noise. In general, a
similar value of $M$ for each DCF bin and a limit of $10\%$ to the
appearance of spurious correlations must be assured in order to get reliable
results. 

Another method frequently used for searching characteristic time scales of
variability is the Discrete Fourier Transform (DFT) spectral analysis for
unevenly sampled data.  We have adopted the implementation of the Lomb
normalized periodogram method (Lomb \cite{lom76}) discussed by Press et al.\
(\cite{pre92}).  The presence of a sinusoidal component of frequency
$\omega_0=2 \pi \nu$ in the dataset is revealed by a large value of the
periodogram $P(\omega)$ at $\omega = \omega_0$.
The significance of the peaks is estimated by the false alarm
probability, i.e.\ the probability that a peak is of height $z$ or higher if
the data are pure noise. It is given by $F=1-(1-{\rm e}^{-z})^M$, where $M$ is the
number of independent frequencies. Since we have scanned frequencies up to the
Nyquist frequency $\nu_c=N/(2T)$ that the $N$ data would have were they evenly
spaced over the period $T$, we have set $M=N$. For very clumpy datasets as we
have when considering the optical light curve, the value of $M$ (and hence the
false alarm probability) may be overestimated, and the significance of the
peaks is consequently underestimated (see also Horne \& Baliunas
\cite{hor86}).

\begin{figure*}  
\epsfxsize=15cm \epsfbox{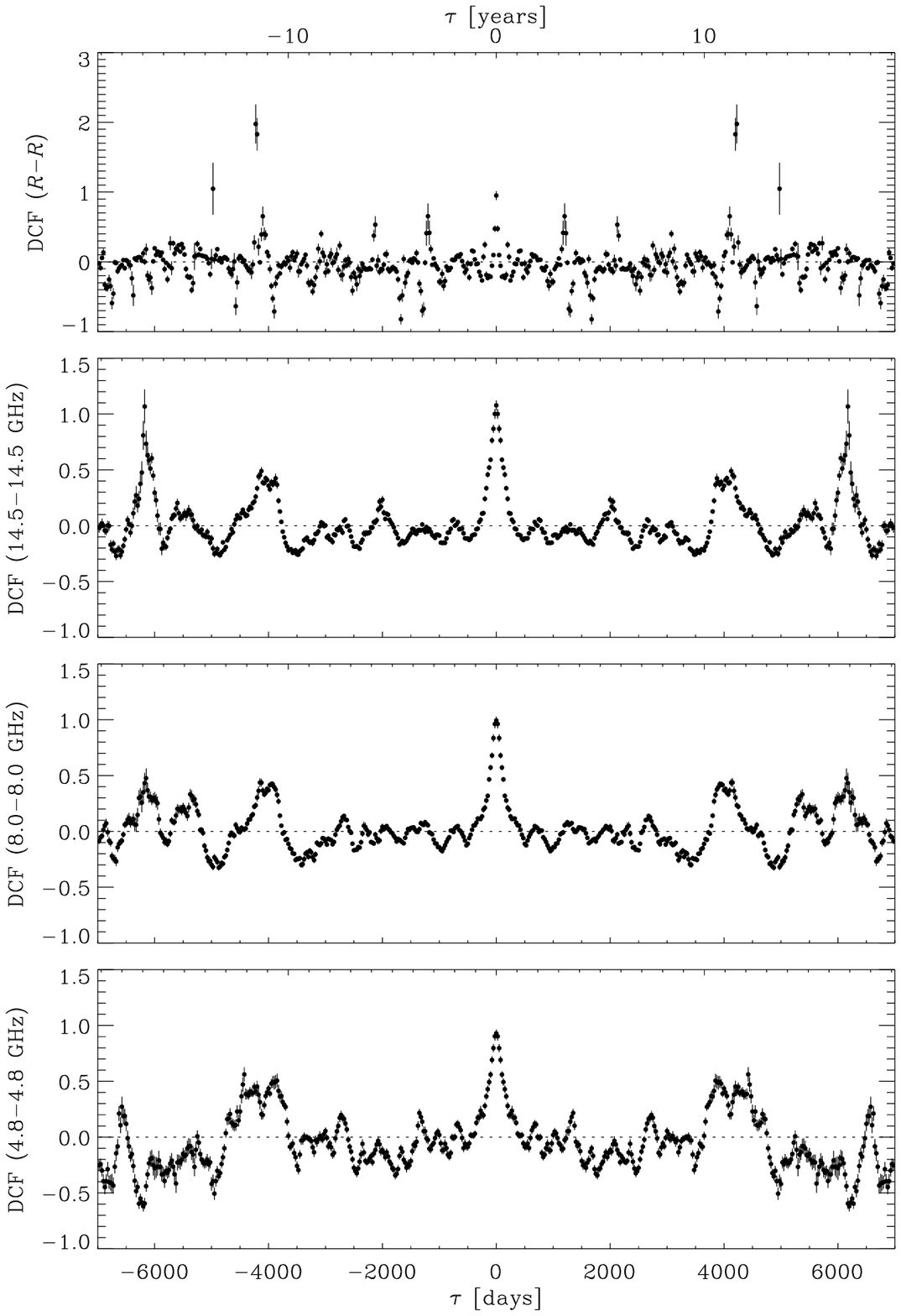}
\vspace{1cm} 
\caption[]{DCF autocorrelations for the optical and UMRAO radio
fluxes shown in Fig.\ \ref{radop}; the data have been
binned over $2$ days, while the DCF was obtained with a bin size of $25$ days}
\label{dcf_auto}      
\end{figure*}  
   
\subsection{Search for characteristic time scales of variability}

\begin{figure*}  
\epsfxsize=15cm \epsfbox{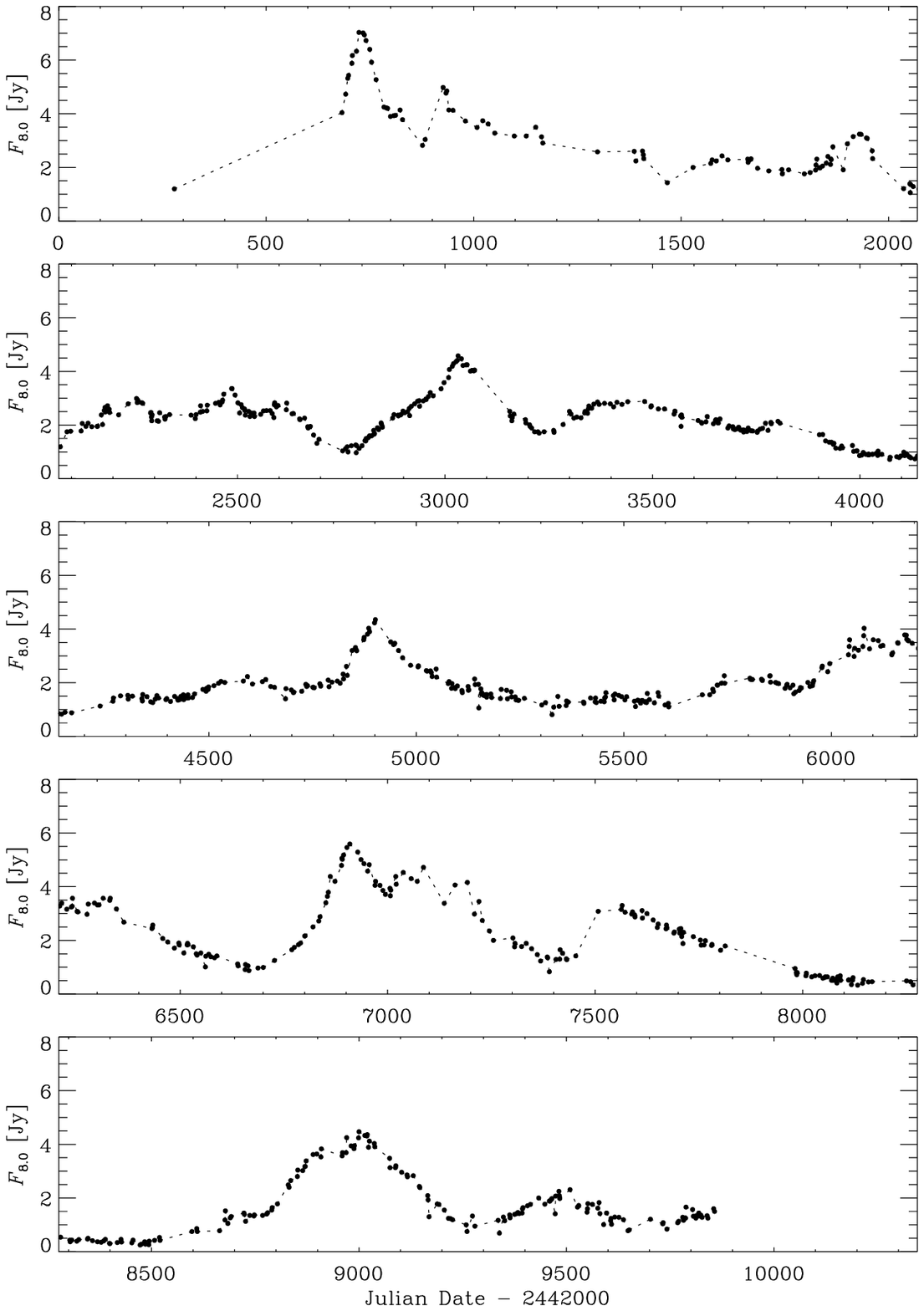} 
\vspace{1cm} 
\caption[]{The light curve of AO 0235+16 at $8.0\rm\,GHz$ folded assuming a period
of $2069$ days}   
\label{folded_8}        
\end{figure*}  

\begin{figure*}  
\epsfxsize=15cm \epsfbox{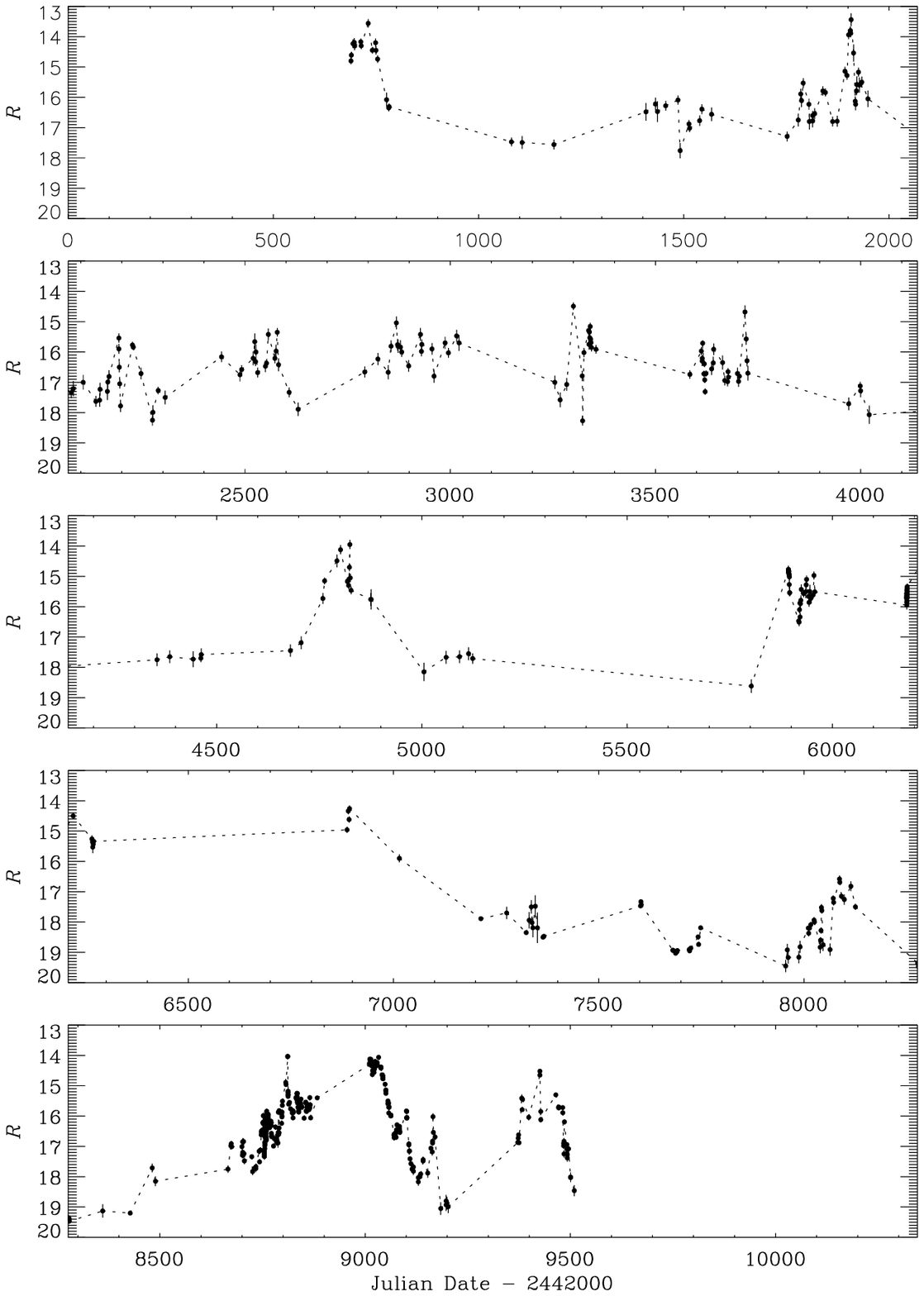}
\vspace{1cm}  
\caption[]{The light curve of AO 0235+16 in the $R$ band folded assuming a
period of $2069$ days}   
\label{folded_r}        
\end{figure*}

The optical DCF autocorrelation is plotted in Fig.\ \ref{dcf_auto} (top panel):
as expected, it is dominated by flicker noise and reveals an important
double-peaked  signal at $\tau \sim 4100$--$4200$ days ($11.2$--$11.5$
years), and not negligible ones at $\tau \sim 1200$ and $\sim 2100$ days
($\sim 3.28$ and $\sim 5.75$ years). These can
thus be regarded as characteristic time scales of optical variability for AO
0235+16. A visual inspection of the optical light curve in Fig.\ \ref{radop} confirms
these features. 

The peak at $\tau \sim 5000$ days in the optical autocorrelation is just one
point affected by a large error, deriving from the correlation between the
1979 and 1992 outbursts.

The fact that the maxima at about $\pm 4200$ days exceed $1$ is
due to the choice of a $2$ day binning on the original dataset; a larger time
interval for data binning (e.g.\ $20$ days) would reduce the importance of these
features with respect to the central maximum, but would force the choice of a
much larger DCF bin ($\sim 100$ days) in order to avoid important spurious
correlations. This in turn would imply missing important details.

Figure \ref{dcf_auto} also presents the DCF autocorrelation for UMRAO
radio fluxes. The peaks are wider than in the optical case,
reflecting the broader outbursts of the radio light curves when compared with
the optical ones. 
An important double-peaked maximum appears at $3900$--$4200$
days ($10.7$--$11.5$ years) at all three frequencies, whose similarity with the
optical one suggests radio-optical correlation. 
Other noticeable peaks are seen at $5300$--$5400$ and $6100$--$6200$ days ($14.5$--$14.8$,
$16.7$--$17.0$ years) in the DCF autocorrelation function for the $8.0\rm\,GHz$ band. At
$14.5\rm\,GHz$ the former appears reduced, while the latter, deriving mainly from the
coupling of the 1975 and 1992 outbursts, is enhanced. At $4.8\rm\,GHz$ both
disappear, but in this case the light curve is less extended in time; in
particular, the 1975 outburst is totally missing.

As for the Mets\"ahovi radio light curves at $22$ and $37\rm\,GHz$, their less dense
sampling and more limited time extension lead to higher spurious effects,
especially for large values of $\tau$. However, their autocorrelation
functions confirm the signal centred at $\tau \sim 4000$ days, and
present noticeable peaks at about $2000$ days, that is the time separation
between the outbursts detected in 1987, 1992, and 1998.

The most interesting point emerging from the above discussion is the $\sim
11.2$ year characteristic time scale of variability, which is common to the
optical and radio fluxes.
By looking at the light
curves in Fig.\ \ref{radop}, one can understand this result by noticing that
at a distance of about $4100$ days the 1975 peak correlates with the 1987 one,
and this latter with the 1998 flare, while the 1982 outburst correlates with
the 1992 one. This means that there seems to be a $\sim 4100$ day
characteristic variability time scale intersecting another $\sim 4100$ day
time scale, which on one side is somehow surprising. On the other side, both
the optical and the Mets\"ahovi radio autocorrelations suggest that there may
be a ``periodicity" of about half the above time scale. Indeed, the $25$ year
time extension of the AO 0235+16 light curves would allow to interpret this
halved time scale in terms of periodicity. The point is to understand why
UMRAO radio autocorrelation functions do not show a strong signal at $\tau \sim
2050$ days. The reason is that this signal is damped by the delay of the 1982
outburst. Indeed, by looking at the $8.0\rm\,GHz$
light curve (Fig.\ \ref{radop}), the best sampled one, one can recognize five
large-amplitude  outbursts peaking at ${\rm JD}-2442000=723.60$, $3031.44$, 
$4901.27$, $6908.74$, and $9000.00$, spaced by $2308$, $1870$, $2007$, and $2091$ days,
respectively. The average period would thus be $2069\pm 184$ days, i.e.\
$5.67\pm 0.50$ years. The $8.0\rm\,GHz$ light curve folded assuming a period of $2069$
days is presented in Fig.\ \ref{folded_8}: the delay of the 1982 outburst is
clearly visible.

Notice that a weak signal at about $2000$ days is actually present in
the $14.5\rm\,GHz$ autocorrelation function because at this frequency
the 1982 outburst was preceded by a kind of pre-outburst that made the flux
reach a high level earlier than the $8.0\rm\,GHz$ one.

We have checked the reliability of the results obtained by the autocorrelation
analysis by means of the Discrete Fourier Transform (DFT) technique for
unevenly sampled data implemented by Press et al.\ (\cite{pre92}). In both the
radio and the optical cases we obtained many signals with significance
levels $F$ better than $0.001$. In particular, the
$2069$ day periodicity previously inferred is confirmed by the DFT analysis 
on UMRAO data.
At $8.0\rm\,GHz$, a clear maximum ($P\sim 79$) is found at frequencies  $\nu \sim
4.8$--$4.9\times 10^{-4} \rm \, day^{-1}$, corresponding to periods of $\sim
2050$--$2080$ days. Other strong signals in the $8.0\rm\,GHz$ data are found at $3.7$,
$2.8$, and $1.8$ years ($P \sim 88$, $46$, $45$, respectively). At $14.5\rm\,GHz$ the
strongest maximum of the Lomb periodogram ($P \sim 81$) is right at $\nu \sim
4.8$--$4.9\times 10^{-4} \rm \, day^{-1}$, followed by the maxima ($P \sim 68$,
$49$, $48$) corresponding to $2.8$, $1.8$, and $3.7$ year periods, confirming the
results obtained for the $8.0\rm\,GHz$ data.  Similar results are also found for the
$4.8\rm\,GHz$ dataset. 

The DFT technique applied to the optical fluxes gives much more
signals, making the spectral analysis rather complex. Surprisingly, the $\sim
2100$ day time scale discovered in the optical autocorrelation function and
obtained by the DFT analysis of UMRAO data gives only a weak signal
(significance level better than $3\%$ only). 
To better visualize the matter, Fig.\ \ref{folded_r} shows the
optical light curve in the $R$ band folded assuming a $2069$ day period.
Notice that such a period would not explain some important outbursts, in
particular the major outburst observed in 1979. 
Other signals obtained by the DFT analysis at $2.8$ and $1.6$--$1.9$ years confirm
the time scales found for the radio fluxes, while there is not a strong
optical signal corresponding to the $3.7$ years in the radio. The
strongest DFT signal corresponds to a $\sim 200$ day time scale, which is the
time separation between the two peaks observed in the 1997--1998 outburst.

\subsection{Optical-radio cross-correlation}

\begin{figure}  
\epsfxsize=9cm \epsfbox{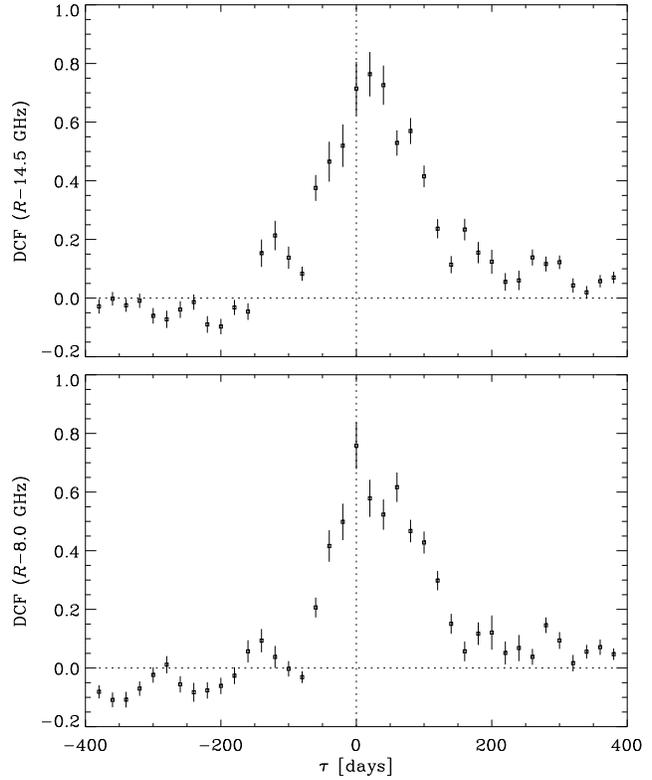} 
\caption[]{Discrete Correlation Function (DCF) between the $R$ data and 
$14.5\rm\,GHz$ (top) or $8.0\rm\,GHz$ (bottom) data of AO 0235+16; all datasets have been
binned over $2$ days, while the DCF bin size is $20$ days} 
\label{dcf_r_radio}       
\end{figure}

The results of the DCF cross-correlation between data in the $R$ band and the
$8.0\rm\,GHz$ ones are shown in Fig.\ \ref{dcf_r_radio} (bottom panel): the
well-defined positive peak at $\tau \sim 0$--$60$ days suggests optical-radio
correlation, with optical variations that can be both simultaneous and leading
the radio ones by a couple of months. The DCF applied to the optical and $14.5\rm\,GHz$ 
datasets leads to a similar result (see Fig.\ \ref{dcf_r_radio}, top
panel): the radio variations appear correlated and delayed of $0$--$50$ days with
respect to the optical ones. These results are in agreement with what was
derived by Clements et al.\ (\cite{cle95}).

\begin{figure}  
\epsfxsize=9cm \epsfbox{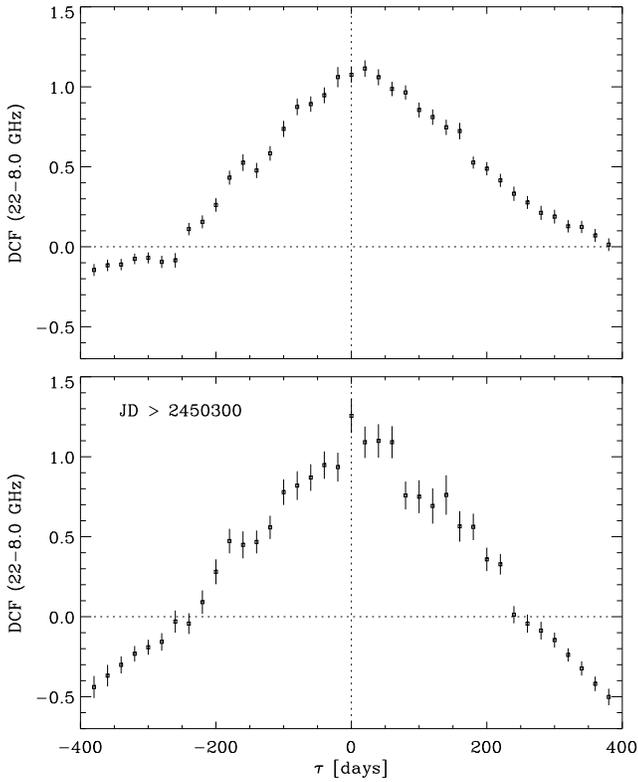} 
\caption[]{Discrete Correlation Function (DCF) between the $22$ and
$8.0\rm\,GHz$ data of AO 0235+16; in the bottom panel only data after ${\rm
JD}=2450300$ have been considered; both datasets have been binned over $2$ days,
while the DCF bin size is $20$ days}  
\label{dcf_22_8}       
\end{figure}

As for the radio-radio correlations, Clements et al.\ (\cite{cle95}) found no
time delay between the $14.5$ and $8.0\rm\,GHz$ datasets and between the $8.0$ and
$4.8\rm\,GHz$ ones. However, as previously discussed, at least during the 1992--1993
and 1998 outbursts, the radio light curves seem to indicate that the flux
variation at the higher frequencies may have led that observed at the lower
ones. Figure \ref{dcf_22_8} shows the results of the DCF cross-correlation
between the $22$ and $8.0\rm\,GHz$ fluxes: in the top panel, where all data have been
considered, the peak is not exactly centred at $\tau=0$. This might suggest
that the $22\rm\,GHz$ fluxes can lead the $8.0\rm\,GHz$ ones by several days. In the bottom
panel of Fig.\ \ref{dcf_22_8} only data after  ${\rm JD}=2450300$ were taken
into account, so that only the 1998 outburst is considered: the delay effect
in this case is enhanced.
This is in agreement with what was observed by O'Dell et al.\ (\cite{ode88})
when analyzing the variability of AO 0235+16 at eight radio frequencies, from
$318\rm\,MHz$ to $14.5\rm\,GHz$. They found that flux-density variations are clearly
correlated, and events occur first at the higher frequencies and propagate
to lower frequencies with decreasing amplitude.

\section{Discussion and conclusions}

Four years of optical and radio monitoring of the BL Lac object AO 0235+16
have confirmed the extreme variability for which this source is well known: 
a range of more than $5\rm\,mag$ was spanned in
the optical band and a total variation of up to a factor $18$ was registered in
the radio fluxes.

Optical spectral changes are not very evident, but our data suggest 
a spectrum steepening when the source gets fainter, a feature that has
already been recognized for other blazars. This behaviour can be interpreted in
terms of radiative losses in the electron population of the emitting jet: when
the source is bright, cooling can be balanced or overcome by acceleration
processes and the resulting spectrum is flatter; when the flux is low,
radiative losses dominate and cause a spectral steepening, since the
higher-energy ultra-relativistic electrons emitting synchrotron radiation
cool faster than the lower-energy ones. However, also a
geometrical interpretation of the kind of that presented in Villata \&
Raiteri (\cite{vil99b}; see also Villata et al.\ \cite{vil00}) is possible:
if a faint state of the source in a given band occurs when the portion of the
curved jet emitting in that band becomes less aligned to the line of sight, 
then a spectral steepening is expected since the higher-frequency
emitting portion of the jet departed first from the line of sight.

Discrete autocorrelation function analysis of the optical and
radio light curves points out a characteristic time scale for the flux
variability of AO 0235+16 of about $11.2$ years. However, a
deeper insight into the optical and radio light curves aided by 
DFT analysis and folded light curves reveals that
major outbursts occur at roughly half the above time
scale, even if one of the events (the 1982 one) appears noticeably delayed. 

Indeed, in the best sampled
$8.0\rm\,GHz$ light curve, there are five large-amplitude outbursts whose
peaks are separated by $2308$, $1870$, $2007$, and $2091$ days, giving an
average ``period" of  $2069 \pm 184$ days, i.e.\ $5.7 \pm 0.5$ years. Taking
into account that the AO 0235+16 redshift is $z=0.94$, the relation
$P=P_{\rm obs}/(1+z)$ implies that the period in the source rest frame is
around $2.8$ years. Other characteristic frequencies resulting from the DFT
analysis of the radio light curves correspond to periods of $1.8$, $2.8$, and $3.7$
years.    
In the optical, the sparse sampling makes the analysis of variability time
scales rather difficult. However, discrete autocorrelation function, DFT, and
folded light curves are compatible with a $\sim 2069$ day periodicity. Such
periodicity would not account for the observed big 1979 optical outburst,
whose radio counterpart was a rather modest flux increase. The optical
autocorrelation analysis puts in evidence also another characteristic time
scale of variability of about $1200$ days ($3.3$ years), which corresponds to the
time separation between the 1975 and 1979 outbursts, while the DFT of the
optical data confirm the $1.8$ and $2.8$ year periods found in the radio data.

A few previous studies investigated the existence of
periodicities in AO 0235+16: Webb et al.\ (\cite{web88}) analyzed the optical
light curve with the Deeming DFT (Deeming \cite{dee75}), and
derived periods of $2.79$, $1.53$, and $1.29$ years. More recently, Webb et al.\
(\cite{web00}) used unequal-interval Fourier transform and CLEAN techniques
and obtained periods of $2.7$ and $1.2$ years for the optical variations. The
time scales of long-term optical base-level fluctuations have been studied by
Smith \& Nair ({\cite{smi95}) for three classes of AGNs; they found a best fit
period of $2.9$ years for AO 0235+16, attributing a moderate confidence to the
estimate. The Jurkevic technique and the autocorrelation function were adopted
by Fan (\cite{fan01}) for the analysis of the optical light curves,
inferring periods of $1.56$, $2.95\pm0.25$, and $5.87\pm1.3$ years. This last
period is highly compatible with what is derived in the present paper. 
Roy et al.\ (\cite{roy00}) made a cross-correlation analysis between optical
and UMRAO radio data and applied Lomb-Scargle periodograms to the $14.5$ and
$8.0\rm\,GHz$ light curves to infer a $\sim 5.8$ year periodicity. Their results are
in agreement with what we find in the present paper.  

If the major outbursts
in AO 0235+16 really occur every $5.7 \pm 0.5$ years, the next one should
peak around February--March 2004,
and a great observational effort should be undertaken since at least summer
2003 to get more information on the details of the flux behaviour in various
bands. Unfortunately, the source is not visible from ground-based optical
observatories during springtime because of solar conjunction. 

On the other
hand, another great effort should be addressed to the theoretical
interpretation of such recurrent events, in particular to envisage possible
mechanisms that can ``disturb" the period.  
Many models have been proposed to explain the $\sim 12$
year period inferred from the optical light curve of another well-known BL Lac
object, i.e.\ OJ 287; most of them foresee the existence of a binary black
hole system  (Sillanp\"a\"a et al.\ \cite{sil88b}; Lehto \& Valtonen
\cite{leh96}; Villata et al.\ \cite{vil98}; Valtaoja et al.\ \cite{val00};
Abraham \cite{abr00}). It would be interesting
to see whether those models can be adapted to explain the quasi-periodic
variability observed in AO 0235+16.    

As for the cross-correlation between the optical and radio data, the analysis
of the light curves is constrained by the sometimes very sparse optical
sampling. However, in general we can say that 
a radio flux increase always corresponds to each major optical flare; 
the vice versa is likely true. Such correlation would suggest that the
same mechanism is responsible for the optical and radio variations. 
However, two different
behaviours seem to be present: optical and radio fluxes are seen sometimes to
reach their maximum values at the same time, while in some cases the higher
frequencies lead the lower ones. This would mean that two different mechanisms
are presumably at work in the long-term variability.  
Outbursts occurring simultaneously in all bands may suggest that the
synchrotron emission from the radio to the optical band is produced by the same
population of relativistic electrons in the same region. But they may also
favour the microlensing scenario, which was already suggested for explaining
the variability of AO 0235+16 (Stickel et al.\ \cite{sti88}; Takalo et al.\
\cite{tak98}; Webb et al.\ \cite{web00}; but see Kayser \cite{kay88} for a
critical discussion).  The presence of several foreground objects in the field
of the source supports this scenario. One item against this hypothesis is that
during the strong flux increase detected in both the optical and radio bands
in 1997, the X-ray and $\gamma$-ray fluxes were not seen in a high state.

On the other hand,
the fact that in the radio and optical domains the variations observed at the
higher frequencies lead those at the lower ones
makes one think to an inhomogeneous jet, where a disturbance
travelling downstream enhances first the emission at the higher frequencies
and then the emission at the lower ones. An alternative view is given by
geometrical models such as the previously mentioned helical
model by Villata \& Raiteri (\cite{vil99b}), if we imagine that the portion of
the jet emitting the higher-energy radiation gets closer to the line of 
sight before that producing the lower-energy flux, as already noted for the
spectral changes.

A different interpretation is required to explain the noticeable intraday
variability shown by AO 0235+16 in both the optical and radio bands.
Indeed, on short time scales, the flux behaviour can be very different with
respect to that exhibited on long time scales. Kraus et al.\ (\cite{kra99})
observed an ``unusual" radio event in October 1992, in which the $20\rm\,cm$ maximum
preceded the maxima at $3.6$ and $6\rm\,cm$, and the variation amplitude was larger at
the lower frequencies. They applied a number of models, both of intrinsic and
of extrinsic nature, and concluded that, in any case, the size of the emitting
region must be very small, implying a Doppler factor of order $100$. Such a
high value was subsequently confirmed by Frey et al.\ (\cite{fre00}) by
analyzing VSOP observations of AO 0235+16.

We found several episodes of very fast intraday variability in the radio light
curves; in some cases we are in the presence of single points that appear to
deviate from the longer trends shown by the data, but there are cases in which
the variation is confirmed by more than one point. Calculation of the
brightness temperature for the fastest variability events ($\Delta F / \Delta
t > 0.3 \, \rm Jy\,day^{-1}$) in the $14.5\rm\,GHz$ light curve (containing data
generally of higher signal-to-noise ratio with respect to the lower
frequencies ones) led to values by far exceeding the Compton limit, implying
Doppler factors in the range $30$--$70$. Indeed, such results must be taken with
caution, since a much better sampling is needed in order to derive reliable
Doppler factors from the radio intraday variability.

In conclusion, all the items pointed out by the present work (such as the $5.7$
year periodicity, delayed flux variations at lower frequencies, high Doppler
factors) would need a further observational effort in the next years, in order
to fix them to more precise results.

\begin{acknowledgements}
We gratefully acknowledge useful suggestions by the referee, Dr.\ K.\ Ghosh.
We thank R.C.\ Hartman for communicating us
the result of the EGRET VP 6311 on AO 0235+16. This research has made use of:
\begin{itemize}
\item the NASA/IPAC Extragalactic Database (NED), which is operated by the 
Jet Propulsion Laboratory, California Institute of Technology, under
contract 
 with the National Aeronautics and Space Administration;
\item data from the University of Michigan Radio Astronomy Observatory,
which 
 is supported by the National Science Foundation and by funds from
the 
 University of Michigan.
\end{itemize}

\end{acknowledgements}

\end{document}